\documentclass{optica-article}
\journal{opticajournal} 
\usepackage{ulem}
\articletype{Research Article}
\usepackage{amsmath}
\newcommand\norm[1]{\lVert#1\rVert}

\usepackage{subcaption}
\usepackage{comment}

\usepackage{cases}
\usepackage{relsize}
\usepackage{graphicx}
\usepackage[linesnumbered,ruled,vlined]{algorithm2e}
\usepackage{amsmath} 
\usepackage{lineno}
\usepackage{xcolor}
 \usepackage{multirow}
\usepackage{array}
\usepackage{makecell}
\usepackage{hhline}
\usepackage{diagbox}

\newcommand{\red}{\textcolor[rgb]{0.00,0.00,0.00}}
\newcommand{\blue}{\textcolor[rgb]{0.00,0.00,0.9}}

\newcounter{algo}
\renewcommand{\thealgo}{\arabic{algo}}

\def\bsk{{\boldsymbol{k}}}

\def\bsm{{\boldsymbol{m}}}

\def\bsA{{\boldsymbol{A}}}

\def\bsD{{\boldsymbol{D}}}
\def\bsE{{\boldsymbol{E}}}

\def\bsH{{\boldsymbol{H}}}

\def\bsR{{\boldsymbol{R}}}

\def\bsV{{\boldsymbol{V}}}

\begin{document}
\title{A Plug-and-Play Algorithm for 3D Video Super-Resolution of Single-Photon LiDAR data}

\author{Alice Ruget,\authormark{1},  Lewis Wilson\authormark{1,2}, Jonathan Leach\authormark{1}, Rachael Tobin\authormark{1},  Aongus McCarthy\authormark{1}, Gerald S. Buller,\authormark{1} Steve McLaughlin,\authormark{1} Abderrahim Halimi\authormark{1*}}

\address{\authormark{1} School of Engineering and Physical Sciences, Heriot-Watt University, Edinburgh, UK}\address{\authormark{2}Imaging Sub-group, STMicroelectronics, Edinburgh EH3 5DA, UK\\}

\email{\authormark{*}a.halimi@hw.ac.uk} 


\begin{abstract*} 

Single-photon avalanche diodes (SPADs) are advanced sensors capable of detecting individual photons and recording their arrival times with picosecond resolution using time-correlated Single-Photon Counting (TCSPC) detection techniques. They are used in various applications, such as LiDAR and low-light imaging. These single-photon cameras can capture high-speed sequences of binary single-photon images, offering great potential for reconstructing 3D environments with high motion dynamics. To complement single-photon data, these cameras are often paired with conventional passive cameras, which capture high-resolution intensity images at a lower frame rate. 
However, 3D reconstruction from SPAD data faces challenges. Aggregating multiple binary measurements improves precision and reduces noise but can cause motion blur in dynamic scenes. Additionally, SPAD arrays often have lower resolution than passive cameras.
To address these issues, we propose a novel computational imaging algorithm to improve the 3D reconstruction of moving scenes from SPAD data by addressing the motion blur and increasing the native spatial resolution. The goal is turning the high-speed SPAD events, recorded at a high frame rate, into non-blurred high-resolution depth images at the frame rate of the passive sensor. We adopt a plug-and-play approach within an optimization scheme alternating between guided video super-resolution of the 3D scene, and precise image realignment using optical flow. 
Experiments on synthetic data show that our method significantly improves image resolution across various signal-to-noise ratios and photon levels. We validate our method using real-world SPAD measurements on three practical situations with dynamic objects. 
First on fast-moving scenes (i.e. fan) in laboratory conditions at short range (3 meters); second very low resolution imaging of people with a consumer-grade SPAD sensor from STMicroelectronics; and finally, high-resolution imaging of people walking outdoors in daylight at a range of 325 meters under eye-safe illumination conditions using a short-wave infrared SPAD camera. These results demonstrate the robustness and versatility of our approach.

\end{abstract*}

\section{Introduction}

Single-photon LiDAR (SPL) is emerging as a technology with diverse applications \cite{Wallace_TVT2020,rapp2020advances,hadfield2023single}, including long-range imaging \cite{pawlikowska2017single,Tobin_OE_2017} for autonomous vehicles and remote environmental sensing, and it is now integrated into smartphones and tablets to improve autofocus and enhance virtual reality \cite{Ruget_SA2023}. It is also used in challenging conditions, such as imaging through obscurants 
\cite{satat2018ICCP,halimi2021robust,tobin2021robust,plosz2023real} or with multiple wavelengths 
\cite{Belmekki_OE23,tachella2019bayesian}. SPL systems employ a time-correlated single-photon counting techniques that offer precise timing measurements (typically picosecond timing) critical for these applications.
The operation of an SPL system involves emitting laser pulses to illuminate a scene and detecting the reflected photons using a single-photon detector (i.e. single-photon avalanche diode SPAD), together with their time-of-flight (i.e. the time taken for photons to travel from the source to the object and back). This process is typically  repeated multiple times and the reflected photons are usually grouped into a histogram of photon counts with respect to the times of flight (ToFs). The histograms generated from SPL measurements are rich in information. They not only allow for precise object detection but also provide accurate estimation of the target's depth and reflectivity.

Single-photon systems face several challenges. In LiDAR imaging for autonomous navigation, capturing moving targets requires short acquisition times to build the histogram of counts to prevent motion blur, and long-range imaging often involves low photon returns, resulting in photon sparse data.
This highlights the need for specialized algorithms to process raw event-based binary data \cite{bocchieri2024scintillation,ma2023burst,ma2020quanta}. Additionally, smartphone applications such as mobile virtual reality demand cost-effective sensors, which currently limit resolution in the cross-sectional plane (i.e., few X-Y pixels) \cite{Ruget_SA2023,Clara_ACM2021}. In these real-world applications, multiple sensors utilizing different technologies are often used simultaneously, requiring advanced data fusion techniques to support decision-making \cite{rapp2020advances,zhang2019deep}. 


Several solutions have been proposed in the literature to address these challenges, which can be grouped into three main categories: model-based, learning-based, and hybrid methods. Model-based approaches rely on statistical or regularization models, and the resulting inference is solved using stochastic simulation methods or optimization algorithms. These methods have been applied to sparse or event-based data \cite{Fast2020det}, long-range imaging \cite{li2020single,pawlikowska2017single}, and LiDAR-passive sensor fusion \cite{halimi2023plug}. While they offer good interpretability, they depend on the definition of appropriate features to represent the data. Learning-based methods, on the other hand, extract important features from training data with ground truth and use these learned features to process new measurements. However, these methods are sensitive to the training data and may require costly retraining if imaging conditions change, such as varying noise levels. They have been used for SPL multimodal fusion and upsampling \cite{lindell2018single,Ruget_OE21,sun2023consistent}, and training with multiple sensors for testing with a single sensor \cite{Ruget_SA2023}. Hybrid methods combine elements of both model-based and learning-based approaches to enhance performance. These include statistical generative models, such as variational auto-encoders (VAEs) and diffusion models, as well as unrolling \cite{zhang2020deep,koo2022bayesian2,Halimi_Arxiv2023} and plug-and-play \cite{tachella2019real,zhang2019deep} approaches, offering improved interpretability alongside strong results. 

Plug-and-Play (PnP) methods have become popular in computational imaging due to their flexible, modular approach \cite{kamilov2023plug}. They typically start with a high-fidelity, physics-based model of the imaging problem and use an optimization framework, where the variables of interest are obtained through an iterative process that alternates between different sub-tasks. Originally, the authors of \cite{Venkatakrishnan_GCSIP_2013} observed that one of these sub-steps could be reduced to a denoising task, which they accomplished by incorporating a state-of-the-art denoiser \cite{Dabov_TIP2007}. This strategy was later expanded to address other sub-tasks by incorporating plugged-in algorithms, such as super-resolution \cite{zhang2019deep}. By combining the strengths of physics-based modeling with advanced deep learning techniques for individual sub-tasks, PnP proves especially effective in scenarios with limited labeled data or when the forward model may change, such as across varying imaging conditions or devices.

In this paper, we present a hybrid computational algorithm for the fusion of multi-modal data to produce high-resolution 3D videos of dynamic scenes. We adopt a plug-and-play framework within an optimization scheme, alternating between motion estimation for aligning binary frames \cite{zach2007duality,wedel2009improved,ipol.2013.26,RanjanCVPR2017} and guided super-resolution using state-of-the-art deep-learning algorithms \cite{sun2023consistent}. This approach enhances robustness to misalignments, leverages high-speed binary data capture for handling fast-moving targets, and minimizes motion blur artifacts, resulting in sharper and more accurate 3D image reconstructions. Our results, demonstrated on both simulated and real-world data, show that the proposed method performs well in challenging scenarios involving noise, motion, and low-resolution data, commonly encountered in real-world applications such as long-range imaging for autonomous navigation and low-resolution imaging with cost-effective sensors.

The main contributions of this work are:
\begin{itemize}
    \item A plug-and-play approach for the alignment and upsampling of single-photon LiDAR (SPL) event data to enhance the reconstruction of 3D images of moving scenes by addressing the motion blur and increasing the spatial resolution of the data beyond the native resolution of the sensor. 
    \item Demonstration of robust performance across various levels of degradation, motion speeds, and noise conditions. 
    \item Validation of the method's effectiveness on real-world experimental data in a range of outdoor and indoor scenarios.
\end{itemize}

The paper is structured as follows: Section 2 formulates the problem and describes the observation model used for SPAD data processing. Section 3 introduces the proposed plug-and-play algorithm for 3D video super-resolution, detailing the core components and optimization steps. In Section 4, we present and analyze the results from both simulated and real-world experiments, demonstrating the effectiveness of our approach across various scenarios. Finally, we present our conclusions in section 5, along with discussing potential directions for future work.

\begin{figure}[h!]
\centering
\begin{subfigure}{0.9\textwidth}
    \includegraphics[width=\textwidth]{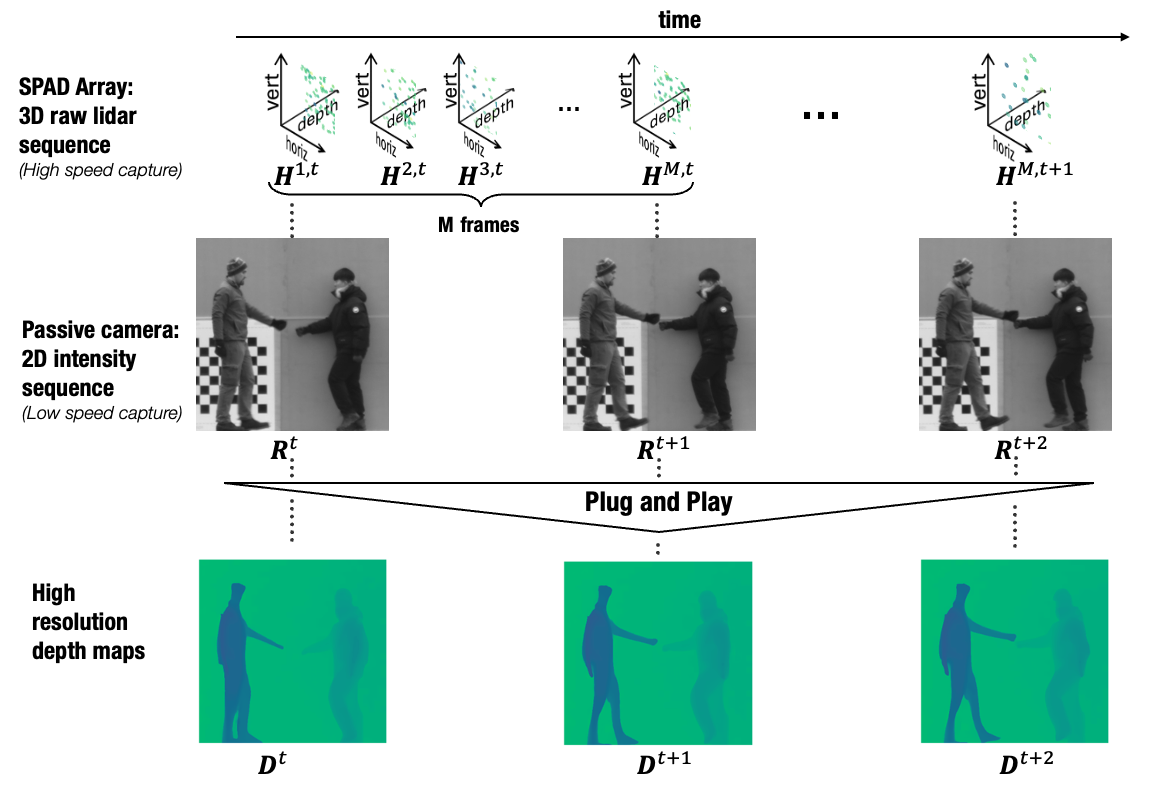}
    \subcaption{\textbf{Overview of the proposed approach.} From left to right: the scene is recorded simultaneously 
    using a LiDAR camera system (based on a SPAD array sensor) and an RGB or passive camera operating at different frame acquisition rates. 
    The proposed plug-and-play algorithm outputs spatial and temporal high-resolution 3D data. We denote $H$ the temporal histograms acquired by the SPAD array sensor, $R$ the high-resolution intensity images and $D$ the high resolution depth maps recovered by our proposed algorithm.  }
    \label{fig:overview}
\end{subfigure}
\par\bigskip
\begin{subfigure}{0.8\textwidth}
\includegraphics[width=\textwidth]{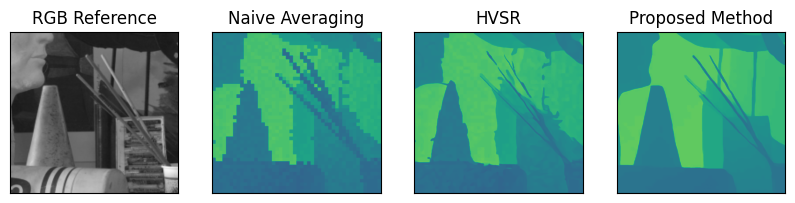}
     \subcaption{\textbf{Results on simulated data:} (1st column)  the reference intensity image; (2nd column) the depth map of the naive averaging method where the binary LiDAR frames are summed without re-alignment; (3rd column) the depth map of the Histogram Video Super-Resolution (HVSR) algorithm \cite{sun2023consistent} on the naive averaged histogram and (4th column) the depth map of the proposed approach. }
    \label{fig:overview2}
\end{subfigure}
\begin{subfigure}{0.08\textwidth}
\raisebox{14mm}{\includegraphics[width=\textwidth]{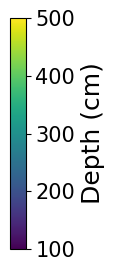}}
\end{subfigure}
\caption{\textbf{Overview of the proposed approach and results.}  }
\label{fig:intro}
\end{figure}

\section{Problem formulation}

In real-world applications, \red{scenes} are often observed with multiple sensors. In this paper, we consider the combination of two camera types - a SPL camera system for acquiring low spatial resolution depth information by providing binary frames at high frame rates, and a passive camera for capturing high spatial resolution intensity images at a lower frame rate. \red{As a result, a large number of binary LiDAR frames are usually collected in the time duration of two consecutive frames from the passive camera. Specifically, we assume that $M$ SPL binary frames are captured between two consecutive passive camera frames (see Fig. \ref{fig:intro}).} 
Note that the passive camera could provide a single intensity channel (gray-scale image), or multiple channels as in the case of a red-green-blue (RGB) camera. There are no restrictions on the technology used for this camera; it could employ SPAD detectors or other imaging technologies.
\red{The goal is to turn the high-speed SPAD events, recorded at a high frame rate, into non-blurred, high-resolution depth images at the frame rate of the passive camera. This involves combining SPAD data with the spatial detail of passive intensity images, while mitigating blur artifacts caused by moving targets.} 
 

SPAD-based LiDAR systems operates by illuminating a target with a short laser pulse (typically a few hundred picoseconds duration) and recording, usually with picosecond resolution, the arrival time of photons reflected back from the scene relative to the laser trigger. 
In its raw form, a SPAD array capture binary frames representing the detection or not of a photon, for each pixel, together with its time-of-flight. In practice, we integrate the binary acquisition of the SPAD recording within a time window of size $M$ frames to generate histogram of photon counts with respect to time-of-flight. Assuming a continuous acquisition of binary frames, the  $N_x \times N_y \times N_{bins}$ histogram at the $t$th time instant can be denoted  $\bsH^t$, where $N_x \times N_y$ represent the number of pixels, and $N_{bins}$ the number of depth bins. Each histogram is constructed using $M$ binary frames, where the $j$th frame is denoted  $\bsH^{j,t}$, with $j \in{1, \cdots, M}$.   
The histogram $\bsH^t$ captures both depth and intensity information about the scene. 
For each pixel $(x,y)$ and time bin $z$, the photon counts $h_{x,y,z}^t$ of the acquired histogram can be expressed as function of the intensity $r_{x,y}$, and the depth $d_{x,y}$ parameters as follows 
\begin{equation}
    h_{x,y,z}^t =\mathcal{P} \left[ r_{x,y}^t  \,  g(z, d_{x,y}^t) + b_{x,y}^t \right],
    \label{image_model}
\end{equation}
where $\mathcal{P}$ denotes a noise operator,  $g$ the impulse response of the system, and $b_{x,y}$ the background level which is assumed constant for all time bins of a given pixel.  Simultaneously, high-resolution intensity or RGB measurements of the scene $\bsR^t$ are recorded, typically using a passive or RGB camera operating at a lower frame rate. The resolution of these images is $f N_x \times f Ny$,  $f $ being the upsampling factor (chosen to be 16 in this paper). If the camera captures RGB measurements, those are converted to intensity images by taking the mean over all channels.




\section{Proposed Method}

\begin{figure}[h!]
\centering

\includegraphics[width=0.85\textwidth]{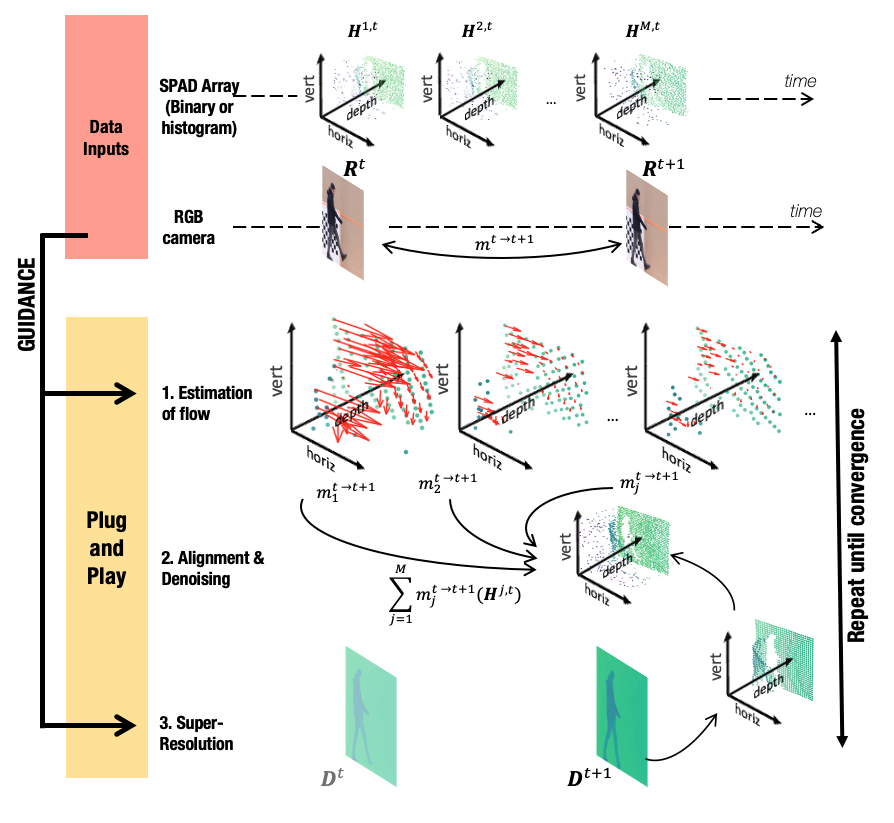}
\label{fig:MPISintel}
\captionsetup{width=0.95\textwidth}
\caption{\textbf{Proposed method.} From top to bottom: The scene is simultaneously recorded using using a single-photon LiDAR camera and a passive camera. The passive camera operates at a lower frame rate than the SPAD sensor, resulting in $M$ SPAD frames $\bsH^{j,t}$ with j $\in [1, M]$ between two consecutive intensity frames $R_{t-1}$ and $R_{t}$.   The proposed plug-and-play approach iteratively performs three steps: first, an alignment step based on the  motion $m^t$ estimated between time $t-1$ and $t$; second, a denoising step merging the aligned frames into a histogram; and third, a video super-resolution step operating on six low-resolution histograms to produce six high-resolution depth images, where $\bsD^t$ represents the $t$th map. }
\label{fig:proposed}
\end{figure}

We propose a novel computational imaging algorithm to improve the 3D reconstruction of moving scenes from SPAD data by addressing the motion blur and increasing the native spatial resolution of the data. We adopt a plug-and-play framework for guided video super-resolution of the 3D scene using an image-guided optical flow for precise realignment.  This algorithm incorporates a state-of-the-art guided video super-resolution algorithm \cite{sun2023consistent} within the iterative process to obtain robust and high-resolution 3D images of the dynamic targets. It processes SPAD binary data in three stages: alignment estimation, denoising, and super-resolution. Figure \ref{fig:proposed}  shows an overview of the approach. 
 \vspace{-0cm}

The goal is to recover successive high-resolution depth maps $\bsD^t, \forall t$ using high-resolution intensity images $\bsR^t, \forall t$, and a series of $M$ temporal histograms acquired between $t - 1$ and $t$, noted $\bsH^{j,t}$ with $j \in [1, M], \forall t$. We assume that each high-resolution depth map results from a high-resolution histogram version $\bsA^t(\bsD^t, \bsR^t)$ obtained at the temporal position of the intensity image. 
The relationship between the series of temporal low-resolution  histograms $\bsH^{j,t}$ and the high-resolution histogram version $\bsA^t$ is defined as follows

\begin{equation}
    \sum_j m_j^{t-1 \rightarrow t} \left( \bsH^{j,t}\right)  =  \left[ \bsA^t(\bsD^t, \bsR^t) \ast \bsk \right]\big\downarrow_s +  \bsE
\end{equation}

where   $\ast \bsk$ represents three-dimensional convolution of $\bsA^t$ with blur kernel $\bsk$ (we considered uniform kernel in the following), $\big\downarrow_s$ denotes the s-fold downsampling operator, and $\bsE$ represents an independent and identically distributed Gaussian noise. 
$m_j^{t-1 \rightarrow t}$ is the motion alignment operator applied to $\bsH^{j,t}$. Assuming a linear movement between $t-1$ and $t$, $m_j^{t-1 \rightarrow t}$, is expressed as a proportion of the global movement $m^{t-1 \rightarrow t}$, expressed as: 
\begin{equation}
    m_j^{t-1 \rightarrow t} = p_j m^{t-1 \rightarrow t}
    \label{eq_Pj}
\end{equation}
where $p_j$ represents the proportion factor specific to each $\bsH^{j,t}$.
The regularized maximum likelihood estimator of the high-resolution histograms $\bsA^t$ can be obtained by minimizing the following cost function:

\begin{equation}
    \min_{\bsD^t, m^{t-1 \rightarrow t}}  \mathlarger{\sum_t} \left|\left| \sum_j m_j^{t-1 \rightarrow t} \left( \bsH^{j,t}\right) -\left[ \bsA^t(\bsD^t, \bsR^t) \ast \bsk \right]\big\downarrow_s\right|\right|^2 + \lambda \Phi \left( \bsR, \bsD \right) + \gamma \Psi \left(\bsR, \bsm  \right) 
    \label{opt_model}
\end{equation}
where $\norm{\sum_j m_j^{t-1 \rightarrow t} \left( \bsH^{j,t}\right) - \left[ \bsA^t(\bsD^t, \bsR^t) \ast \bsk \right]\big\downarrow_s}^2$ is the data fidelity term,   $\Phi \left( \bsR^t, \bsD^t \right)$ the regularization term depending on $\bsR^t, \forall t$ the 2D reflectivity map and $\bsD^t, \forall t$ the 2D depth map, $\Psi \left(\bsR, \bsm  \right)$ is the motion regularization term,  and $\lambda,  \gamma$ the trade-off parameters.  
We consider the half-quadratic splitting (HQS) algorithm which tackles \eqref{opt_model} by introducing an auxiliary variable $\bsV^t$, leading to the following cost function:
\begin{equation}
    \min_{\bsD^t, \bsV^t, m^{t-1 \rightarrow t}}  \mathlarger{\sum_t}\left|\left|\sum_j m_j^{t-1 \rightarrow t} \left( \bsH^{j,t}\right)  - \bsV^t\right|\right|^2 + \mu \left|\left|\bsV^t - \left[ \bsA^t(\bsD^t, \bsR^t) \ast \bsk \right]\big\downarrow_s\right|\right|^2 + \lambda \Phi \left( \bsR, \bsD \right)  + \gamma \Psi \left(\bsR, \bsm  \right) 
\end{equation}
with $\mu$ the penalty term. 

Such problems can be addressed by iteratively solving sub-problems for $\bsD^t,  \bsV^t, \bsm$,  and is  split into three intermediate optimization problem steps:  
{\scriptsize
\begin{numcases}{}
  \min_{m^{t-1 \rightarrow t}}  
  & $\displaystyle\sum_t \left\| \bsV^t - \sum_j p_j m^{t-1 \rightarrow t} \left( \bsH^{j,t} \right) \right\|^2 + \gamma \Psi \left(\bsR, \bsm \right)$ \label{flow} \\
  \min_{\bsV^t} 
  & $\displaystyle\sum_t \left\|\sum_j m_j^{t-1 \rightarrow t} \left( \bsH^{j,t} \right) - \bsV^t \right\|^2 + \mu \left\| \bsV^t - \left[ \bsA^t(\bsD^t, \bsR^t) \ast \bsk \right]\big\downarrow_s \right\|^2$ \label{data_term} \\   
  \min_{\bsD^t}  
  & $\displaystyle\sum_t \mu \left\|\bsV^t - \left[ \bsA^t(\bsD^t, \bsR^t) \ast \bsk \right]\big\downarrow_s \right\|^2 + \lambda \Phi \left( \bsR, \bsD \right)$ \label{prior_term}
\end{numcases}}  
Equation \eqref{flow} represents motion estimation to align the binary frames, \eqref{data_term} represents a guided denoising problem, and \eqref{prior_term} represents the guided super-resolution step. The following paragraphs discuss each of these estimation problems.   

\subsection{Estimation of flow}

Problem \eqref{flow} aims at estimating the scene motion between times $t-1$ to $t$. This motion is estimated using a plug-and-play approach with optical flow techniques.  The motion operators $m^{t-1 \rightarrow t}$  in the horizontal and vertical directions can be obtained with 
several optical flow algorithms, in this paper we considered  the TV-L1 optical flow algorithm \cite{zach2007duality,wedel2009improved,ipol.2013.26} (note that other strategies could be considered such as SPyNet \cite{RanjanCVPR2017}). This algorithm is applied to two consecutive high-resolution intensity images: $\bsR^{t-1}$ at time $t-1$ and $\bsR^{t}$ at time $t$ to estimate the motion between these two time points. The optical flow is applied to each $\bsH^{j,t}$ for j $\in [1, M]$ following \eqref{eq_Pj} with $p_j = \dfrac{M-j}{M-1}$, so that all the frames are referenced to the intensity image $\bsR^{t}$, at time t. 

The motion in the depth direction is determined using the high-resolution  depth maps, $\bsD^{t-1}$ and $\bsD^{t}$, which are produced by the super-resolution step. $\bsD^{t-1}$ is  realigned in the x and y directions using the TV-L1 optical flow algorithm to match the data $D_{t}$ at time $t$. The displacement map is calculated as the difference between $\bsD^{t}$ and the realigned $\bsD^{t-1}$, and is applied to the temporal histograms $\bsH^{j,t}$  with an alignment factor of $p_j$.

\subsection{Guided denoising: alignment and averaging}
Equation \eqref{data_term} highlights a guided denoising problem where the observation are the aligned binary frames, and the guidance is the downsampled well-aligned high-resolution histogram obtained using the high-resolution depth map $\bsD^t$. The two terms appear within an $\ell_2$ norm leading to the closed-form expression :
\begin{equation}
\bsV^t = \frac{ \mathlarger{\sum_j}{ m_j^{t-1 \rightarrow t} \left( \bsH^{j,t}\right)}   + \mu  \left[ \bsA^t(\bsD^t, \bsR^t) \ast \bsk \right]\big\downarrow_s}{\mu + 1}
\end{equation}

\subsection{Guided video super-resolution}
\label{section_HVSR}
Equation \eqref{prior_term} represents a guided super-resolution problem aiming at estimating a high-resolution depth map $\bsD^t$ (or equivalently a high-resolution histogram $\bsA^t$) from low-resolution histograms  $\bsV^t$ under the guidance of high-resolution reflectivity maps $\bsR^t$. Several algorithms were designed to perform this task with deep learning strategies showing good performance \cite{lindell2018single,halimi2023plug,plosz2023fast}.    Adopting a PnP framework, we solve this sub-problem by plugging the state-of-the-art Histogram Video Super-Resolution (HVSR) deep-learning algorithm  \cite{sun2023consistent}. The HVSR algorithm is selected because of it operates on a sequence of histogram data  hence accounting for the scene temporal correlations and improving the quality of the 3D video. More precisely, HVSR is built to take as input a sequence of six low-resolution  histograms of size $N_x \times N_y \times N_{bins}$ and the corresponding high-resolution intensity images of size $f N_x \times f Ny$, with a fixed super-resolution factor $f = 16$. Each histogram and corresponding intensity must be aligned in space and time. 
HVSR outputs six high-resolution depth images each of shape $f N_x \times f Ny$. Each high-resolution map $\bsD^t$ and reflectivity image $\bsR^t$ will then be used to compute a high-resolution histogram $\bsA^t$ using the image model in \eqref{image_model}.

\begin{algorithm}
\caption{Proposed algorithm for guided video super-resolution (we used $T=6$)}
\label{pseudocode_version}
\textbf{\underline{Initialization:}}

\For{$t \in [1, T] $}{
    1. Estimate motion in 2D using $\bsR^{t-1}$ and $\bsR^{t}$ \tcp*{Section 3.1}
    2. Align in 2D and compute the histogram  $ \mathlarger{\sum_j}{ m_j^{t-1 \rightarrow t} \left( \bsH^{j,t}\right)} $ \tcp*{Section 3.2}
    3. Estimate the auxiliary variable $\bsV^t$ using Equation (9) \tcp*{Section 3.2}   
}
    4. Perform video SR using HVSR on $\bsV^t$ with guidance $\bsR^t$ to obtain $\bsD^t$, for $t \in [1, T]$ \tcp*{Section 3.3} 
    
\textbf{\underline{Plug-and-play algorithm:}}

\While{stopping criteria is not met}{
    \For{$t \in [1, T]$}{
        1. Estimate 3D motion using $\bsR^{t-1}$, $\bsR^t$, $\bsD^{t-1}$, and $\bsD^t$ \tcp*{Section 3.1}
        2. Apply guided denoising to obtain $\bsV^t$ \tcp*{Section 3.2} 
    }
    3. Perform video SR using HVSR on $\bsV^t$ with guidance $\bsR^t$ to obtain $\bsD^t$, for $t \in [1, T]$  \tcp*{Section 3.3}
}
\end{algorithm}

\subsection{Stopping criteria}

Algo. \ref{pseudocode_version} summarizes the main steps of the proposed algorithm. 
This iterative process requires the definition of stopping criteria. In this paper, we define two criteria for algorithm termination, with the process stopping when either is met.
The first criterion assesses the updated depth parameter values and stops the algorithm when the  Root Mean Square Error (RMSE) between the current and previous iterations is less than a threshold (fixed to 5 cm). The second criterion is based on a maximum number of iterations, to prevent excessive computation and ensure efficiency.


\section{Experiments }

\subsection{Evaluation metrics and comparison algorithms}
We quantify the performance of the methods on simulated data by comparing the reconstruction with the ground truth depth using two different metrics. 
\begin{itemize}
\item  \textit{Root Mean Square Error (RMSE)}: The RMSE is computed as: RMSE= $\sqrt{\frac{1}{N} \norm{ D - D^{\textrm{ref}}}^2}$, $D$ being the high-resolution depth map predicted by our method and $D^{\textrm{ref}}$ the ground truth.
\item \textit{Percentage of Correct Pixels}: This metric calculates the percentage of pixels where the absolute error is below a given threshold (\(T_h = 5\) cm and \(T_h = 3\) cm). The percentage is computed as:
\[
\text{percentage} = \frac{\sum_{i=1}^{N} \left[ |D_{x,y} - D^{\textrm{ref}}_{x,y}| < T_h \right]}{N} \times 100.
\]
\end{itemize}

We compare the results of our method with the following algorithms.
\begin{itemize}
    \item \textit{Naive Averaging} Depth is computed by summing successive binary frames into a low-resolution histogram and then estimating the depth parameter using the maximum likelihood strategy (i.e., the center-of-mass of the main peak as in \cite{gyongy2020high}).
     \item \textit{HVSR} \cite{sun2023consistent} Depth is obtained by applying the neural network HVSR to the low-resolution histogram generated using naive averaging, along with the corresponding high-resolution intensity or RGB image. 
\end{itemize}

\begin{figure}[h!]
\centering
\begin{subfigure}{0.45\textwidth}
    \includegraphics[width=\textwidth]{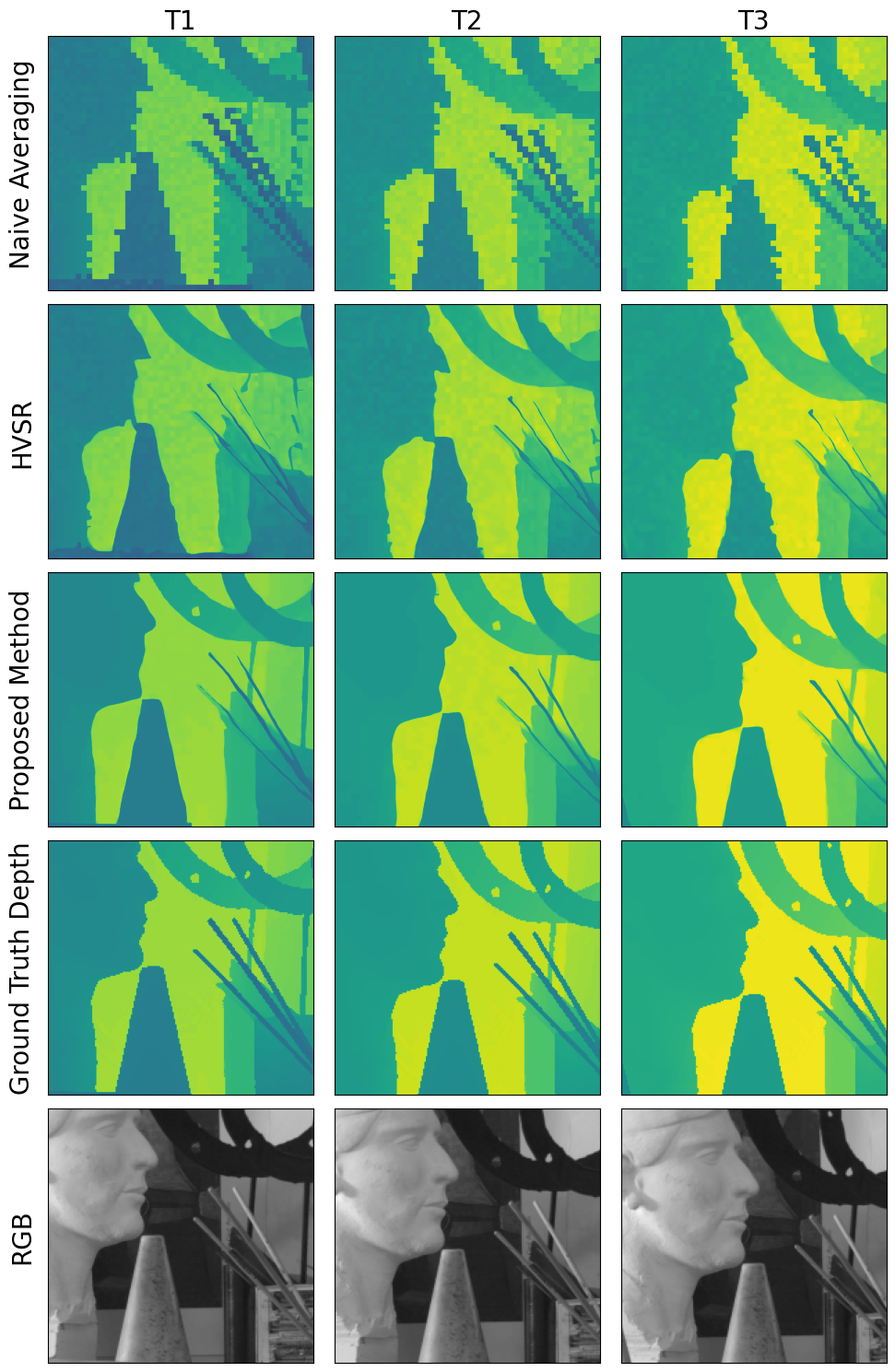}
    \subcaption{Small movement speed}  
\end{subfigure}
\hfill
\begin{subfigure}{0.45\textwidth}
    \includegraphics[width=\textwidth]{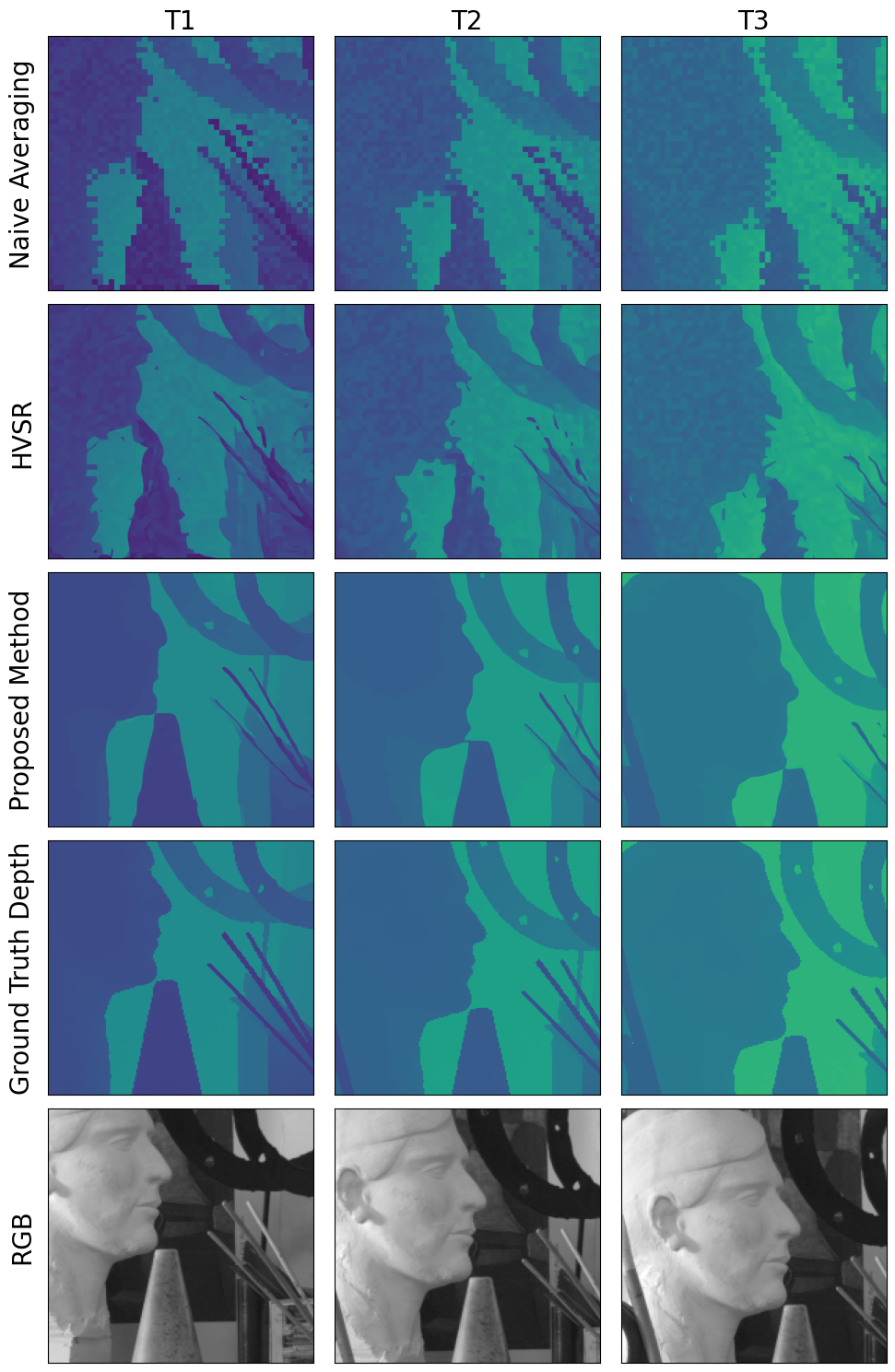}
    \subcaption{High movement speed}  
\end{subfigure}
\hfill
\begin{subfigure}{0.08\textwidth}
    \raisebox{3mm}{\includegraphics[width=\textwidth]{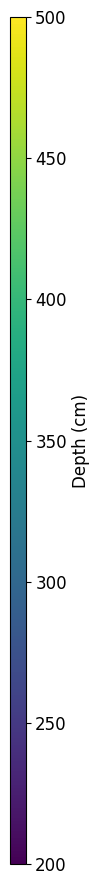}}
\end{subfigure}
\captionsetup{width=0.95\textwidth}
\caption{\textbf{Results on simulated data for different movement speeds.} \red{To simulate motion, we artificially shift the entire image in all xyz directions across frames.}
(a) presents the results for a small movement speed of 0.1 pixels per binary frame (equivalent to 0.21 cm) along the x and y axes, leading to a total displacement of 21 cm between two intensity images. In the depth direction, the movement speed is 0.1 bins per binary frame, corresponding to a depth change of 5 cm between two intensity images. A video representation of this scenario is provided in \blue{Visualisation 1}. (b) shows the results for double the movement speed, with a displacement of 42 cm in the x and y axes, and 10 cm in the z-axis between two intensity images. The simulations were conducted with an SBR of 16 and a ppp of 64. }
\label{results_simulated}
\end{figure}

\subsection{Evaluation on simulated dataset}

\begin{figure}[h!]
\centering
\begin{subfigure}{0.24\textheight}
\includegraphics[width=0.8\textwidth]{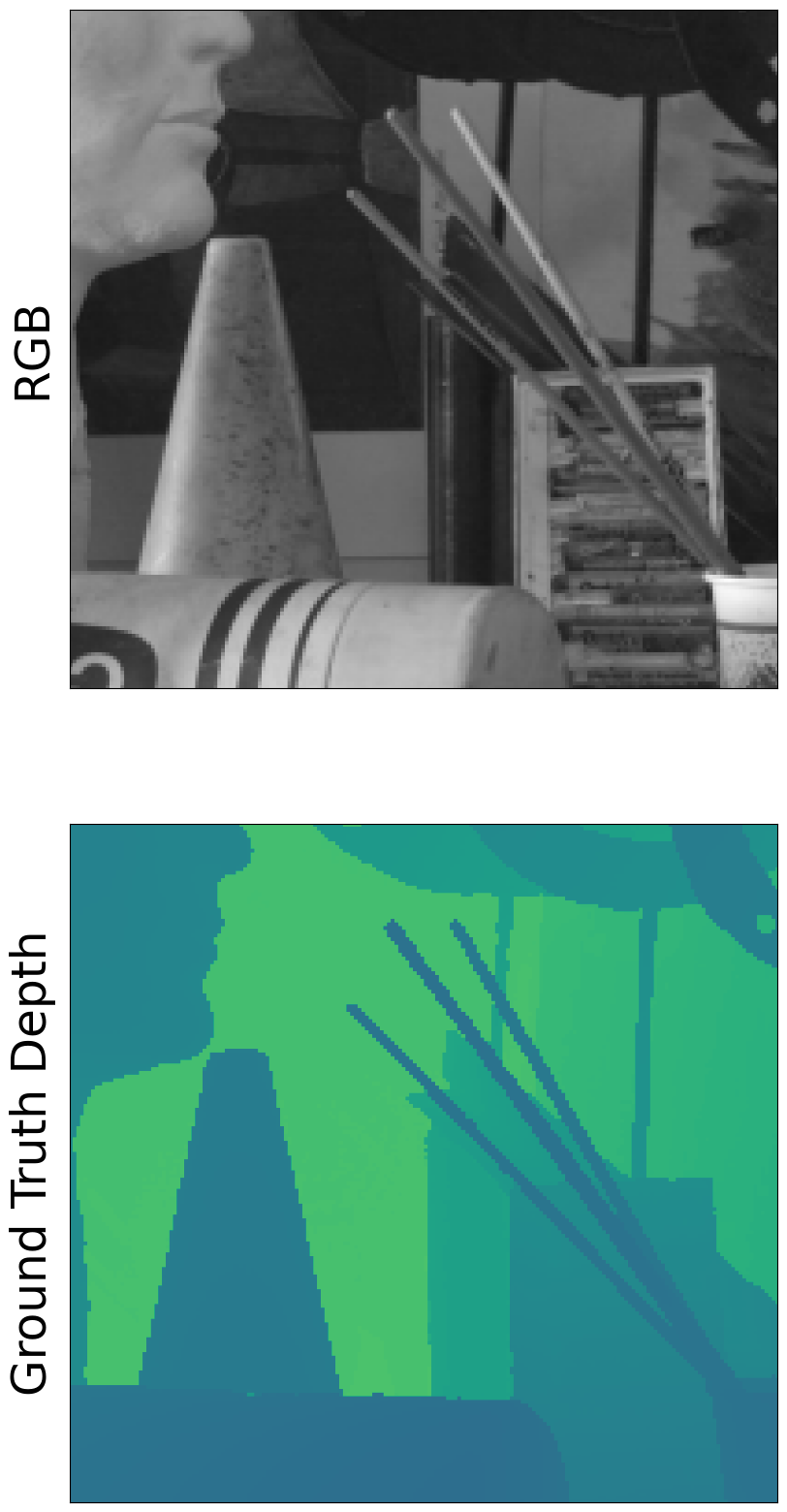}
\end{subfigure}
\begin{subfigure}{0.56\textwidth}
\includegraphics[width=\textwidth]{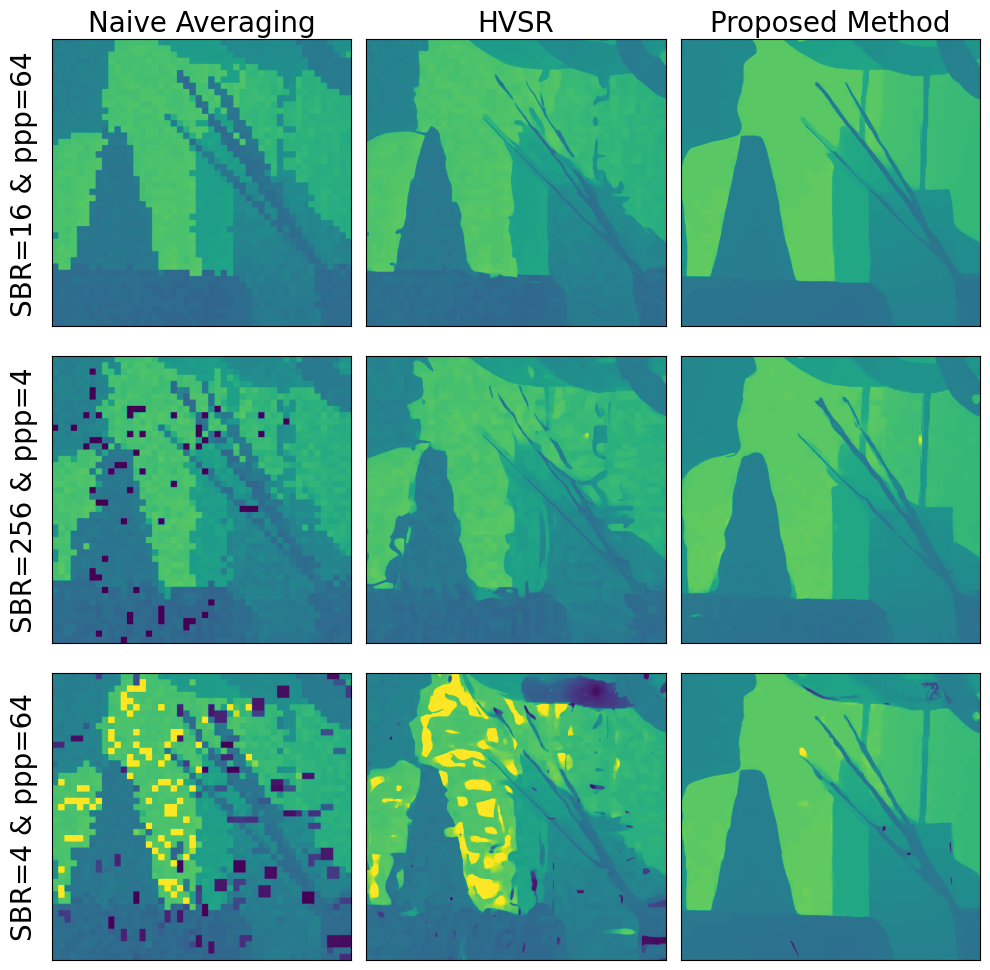}
\end{subfigure}
\hfill
\begin{subfigure}{0.07\textwidth}
    \includegraphics[width=\textwidth]{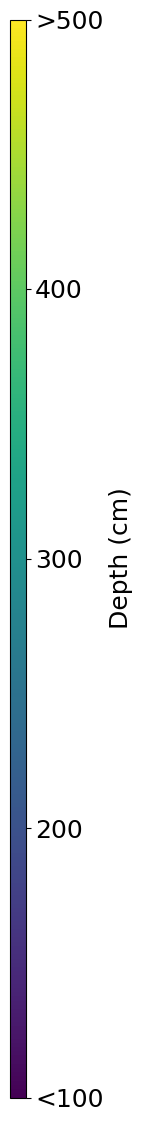}
\end{subfigure}
\captionsetup{width=0.95\textwidth}
\caption{\textbf{Results on simulated data for different noise conditions.} On the left, we display the intensity image alongside the high-resolution ground truth depth map used as a reference. On the right, we present the results for three different noise conditions: the top row corresponds to good noise levels with an SBR of 16 and 64 photons per pixel (ppp); the middle row shows the results for a sparse regime with SBR = 256 and ppp = 4; and the last row displays results for a noisy situation with SBR = 4 and ppp = 64. The simulations are performed with a movement speed that results in a displacement of 21 cm in the x and y directions, and 5 cm in depth, between consecutive intensity images.}
\label{results_noise}
\end{figure}

\begin{figure}[h!]
\centering
\includegraphics[width=\textwidth,height=5cm]{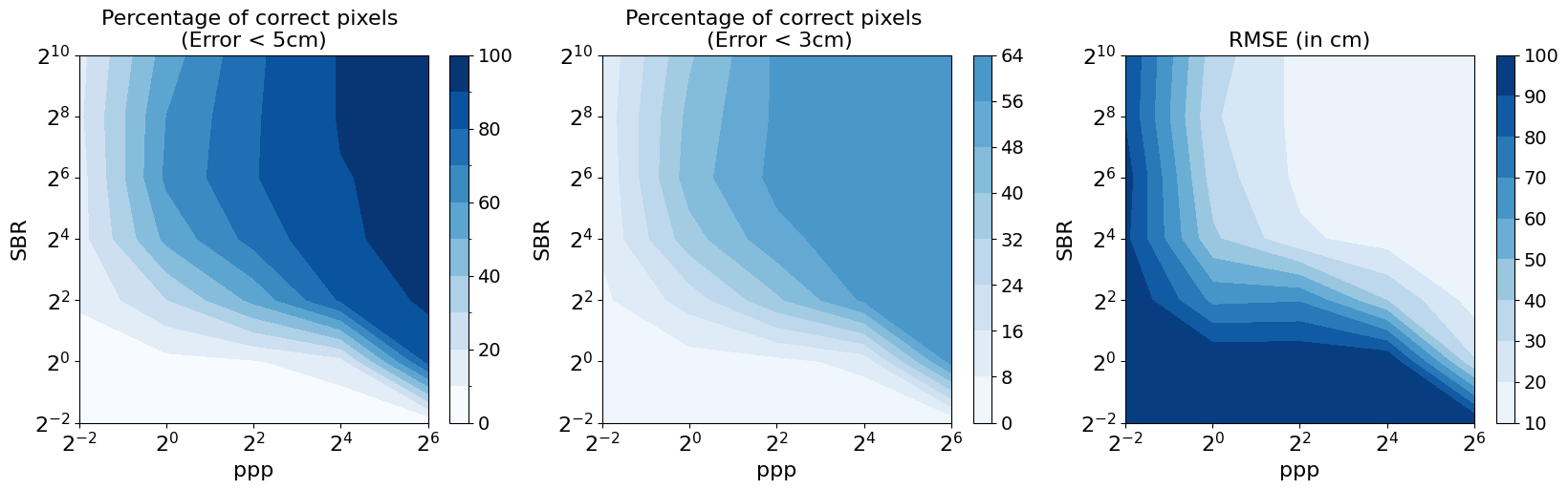}
\captionsetup{width=0.95\textwidth}
\caption{\textbf{Evaluation metrics for simulated data across different noise levels.} From left to right, we present plots of the evaluation metrics across different noise levels. The y-axis represents the SBR, ranging from $2^2$ to $2^{10}$, and on the x-axis, the ppp varies from $2^{-1}$ to $2^6$. The first two plots show the percentage of correct pixels for thresholds of 5 cm and 3 cm. The third plot displays the Root Mean Square Error (RMSE) in centimeters. The simulations are conducted with a movement speed corresponding to a displacement of 21 cm along the x and y axes, and 5 cm in depth between two RGB images.
}
\label{rmse_plot}
\end{figure}

We choose the widely-used stereo Middlebury dataset \cite{scharstein2007learning, hirschmuller2007evaluation} to quantitatively evaluate different methods.
We simulate realistic SPAD array measurements from the high resolution image and depth map of the art scene of the Middlebury dataset. We generate synthetic data following the model in \eqref{image_model} with T = 100 bins where each bin represents 5.52cm, and assuming a Gaussian impulse response $g$ with standard deviation of 4cm. We consider different levels of signal-to-background ratio ($\textrm{SBR} = \dfrac{\sum_{x,y} r_{x,y}}{\sum_{x,y} b_{x,y}}$), average number of photons per pixel ($\textrm{ppp} = \dfrac{1}{N_xN_y} \sum_z h_{x,y,z}^t$) and speed of movement. The $896 \times 1024$ depth measurements are down-sampled by a factor of $16$ to $56 \times 64$ pixels. We simulate movement in the dataset by linearly shifting the entire image in all $xyz$ directions. \red{This does not represent realistic object movement but is designed to test algorithm performance under controlled 3D displacements}
The speed of movement in $x$ and $y$ is expressed in number of pixels per binary frames, and in $z$ using the number of time bins per binary frame. We choose the number of LiDAR frames obtained between two high-resolution intensity frames to be equal to $M = 100$. 
We experimentally analyze the results for different imaging speeds, SBR and ppp level, and identify the imaging regimes where it provides significant benefits.
Figure \ref{results_simulated} presents the reconstruction results for various methods. In Figure \ref{results_simulated} (a), the results are shown for a displacement of 0.1 pixels per binary frame in the x and y directions, and 0.1 time bins per binary frame in depth, resulting in a total displacement of 21 cm in the xy-plane and 5 cm in the z-direction between two intensity images. Figure \ref{results_simulated} (b) depicts the results for double the movement speed, corresponding to a displacement of 42 cm in the x and y axes and 10 cm in the z-axis between two intensity images. The simulations were conducted assuming good noise conditions with an SBR of 16 and a ppp of 64. In both cases, our method produces more accurate reconstructions with sharper edges. Figure \ref{results_noise} illustrates the reconstruction performance of different methods (naive averaging, HVSR \cite{sun2023consistent}, and our proposed method) under varying noise levels. Unlike HVSR and naive averaging, our method demonstrates robustness across different noise levels. Figure \ref{rmse_plot} shows the evaluation metrics across varying noise conditions. The plot includes the RMSE values and the percentage of correct pixels for thresholds of 3 cm and 5 cm, comparing the ground truth with the reconstruction obtained using our method. The metrics are plotted against different SBR and ppp values. The results indicate that our proposed method performs consistently well across a wide range of SBR and ppp levels. The method processes six frames simultaneously over ten iterations, with a total processing time of 56 seconds per iteration. Each iteration consists of two steps: the motion estimation and alignement process takes 39 seconds, and the super-resolution network HVSR \cite{sun2023consistent}, which runs on an NVIDIA RTX 6000 GPU, takes 17 seconds.

\subsection{Evaluation on real 3D SPAD videos }
The proposed algorithm is validated on real data while considering three practical situations with dynamic objects using different SPAD systems. The first two cases are conducted indoor under lab conditions while the last one is performed in outdoor conditions. The first scenario studies fast moving objects, the second considers very low resolution imaging with affordable SPAD sensors (i.e., pixel sparsity), and the last case considers sparse-photon data when acquiring binary frames under outdoor conditions. Table \ref{table_systems} summarizes the experimental setup details and specifications of the different SPAD systems used in each scenario. 

\begin{table}[h!]
\centering
\renewcommand{\arraystretch}{0.9}
\setlength{\tabcolsep}{1pt}
\begin{tabular}{p{4cm}|c|c|c}
\hhline{====}
 \multicolumn{4}{c}{\textbf{LiDAR Systems}} \\
\hhline{====}
 & \textbf{QuantIC4x4} & \makecell{\textbf{STMicroelectronics} \\ \textbf{VL53L8 SPAD}} 
 & \makecell{ \textbf{Princeton Lightwave Kestrel}} \\
\hhline{====}
Pixels & 64 $\times$ 32 & 8 $\times$ 8 & 32 $\times$ 32 \\
\hline
Frame rate & $\sim$1000 Hz & 30 Hz & 150,400 Hz \\
\hline
Bin width & 700 ps & 250 ps & 250 ps \\
\hline
Number of bins used & 16 & 144 & 100 \\
\hline
Illumination wavelength & 670 nm (VIS) & 940 nm (NIR) & 1550 nm (SWIR) \\
\hhline{====}
\multicolumn{4}{c}{\textbf{Measurement scene}} \\
\hhline{====}
Scene description & Rotating axial fan & Moving person & Moving people \\
 & (1000 rpm) & & \\
\hline
Location/lighting & Indoors, ambient & Indoors, ambient & Outdoors, broad daylight \\
\hline
Standoff distance & 3 m & 3 m & 325 m \\
\hline
Field of view & [0.6 $\times$ 1.2 m] & [2.4 $\times$ 2.4 m] & [2.1 $\times$ 2.1 m] \\
\hhline{====}
\multicolumn{4}{c}{\textbf{Intensity image acquisition}} \\
\hhline{====}
Camera & QuantIC4x4 sensor & Kinect camera & Canon EOS 90D DSLR \\
\hline
Pixels (LiDAR FoV) & 256 $\times$ 128 & 128 $\times$ 128 & 500 $\times$ 500 \\
\hline
Frame rate & $\sim$1000 Hz & 7.5 Hz & 25 Hz \\
\hhline{====}
\end{tabular}
\caption{Specifications and measurement setup of the considered SPAD Systems.}
\label{table_systems}
\end{table}


\vspace{-0.2cm}
\subsubsection{Reconstruction of a fast-rotating axial fan in an indoor environment}
\begin{figure}[h!]
\centering
\begin{subfigure}{0.45\textwidth}
    \includegraphics[height=0.85\textwidth]{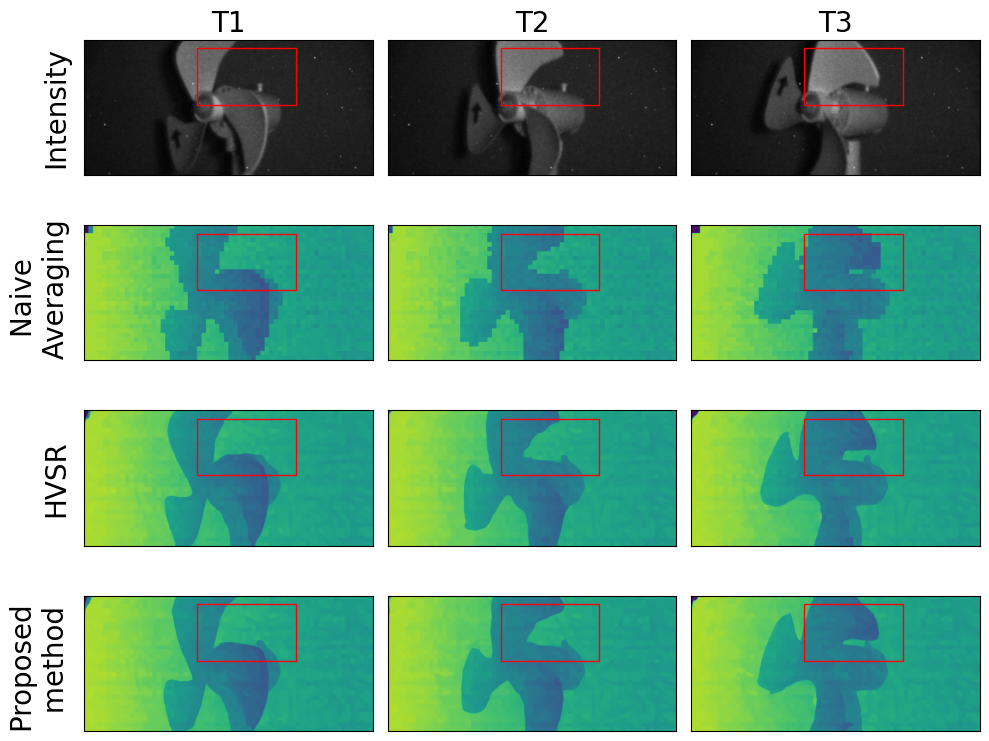}
    \subcaption{Full image results. }
\end{subfigure}
\hfill
\begin{subfigure}{0.45\textwidth}
    \includegraphics[height=0.85\textwidth]{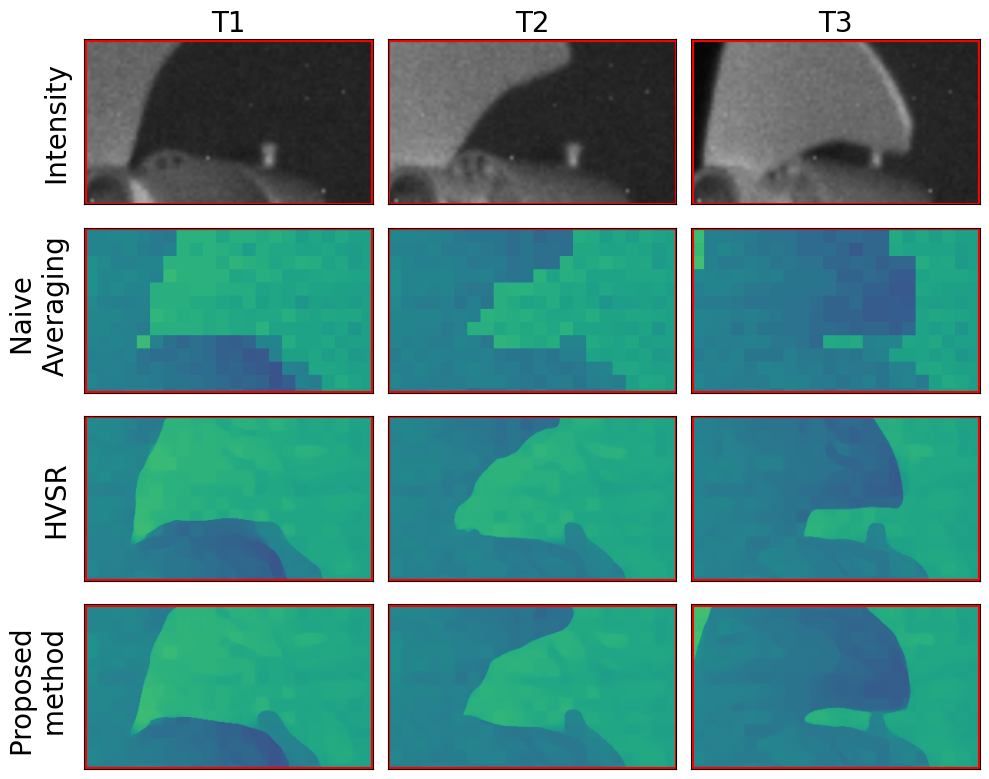}
     \subcaption{Closeup views.}
\end{subfigure}
\hfill
\begin{subfigure}{0.065\textwidth}
    \raisebox{3mm}{\includegraphics[width=\textwidth]{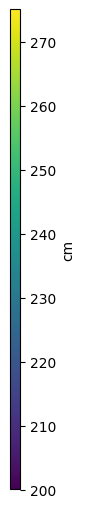}}
\end{subfigure}
\captionsetup{width=0.95\textwidth}
\caption{ \textbf{Intensity and depth imaging of a 3 bladed rotating axial fan using the QuantIC4x4 LiDAR system.} Each column represents a different time frame, and the right figure shows a zoomed-in of the left figure. First row displays the reflectivity image from the SPAD array; Second row   is the reconstruction of the depth using naive averaging of the binary frames; Third row is the reconstruction of the depth using HVSR \cite{sun2023consistent} applied to the image of second row; and fourth row is the reconstruction of the proposed method. A video representation of this scenario is provided in \blue{Visualisation 2}. }
\label{fig:fan_data}
\end{figure}

We used the Quantic4x4 SPAD array sensor,  that generates a histograms of counts on-chip and operates in a hybrid acquisition mode \cite{hutchings2019reconfigurable,henderson20195,gyongy2020high}. This hybrid mode alternates between two measurement modes: a high-resolution intensity measurement, and a low-resolution  histogram of photon counts. The characteristics of the camera are detailed in Table \ref{table_systems}. The two modes alternate, meaning that the histogram and intensity data are not captured simultaneously. The intensity image has a native resolution that is four times higher than the resolution of the depth data. To meet the requirements of the HVSR algorithm, which needs a super-resolution factor of 16, we upsample the high-resolution intensity data by a factor of four using the nearest neighbor algorithm. The scene pictures a fast-rotating axial fan, which is a non linear movement, unlike the linear motion modeled in the simulated data. In this specific case, $M$ is equal to one, meaning there is only one histogram captured between two intensity images.
Figure \ref{fig:fan_data} presents the reconstructions using different methods: naive averaging, HVSR, and our proposed method. The results show that our method produces more accurate images with sharper edges.
The method processes six frames simultaneously over two iterations, with a processing time of 58 seconds per iteration. Each iteration consists of two steps: the motion estimation and alignment process which accounts for 38 seconds, and the super-resolution network HVSR \cite{sun2023consistent}, which runs   on an NVIDIA RTX 6000 GPU, takes 20 seconds.

\begin{figure}[h!]
\centering
\includegraphics[width=0.8\textwidth]{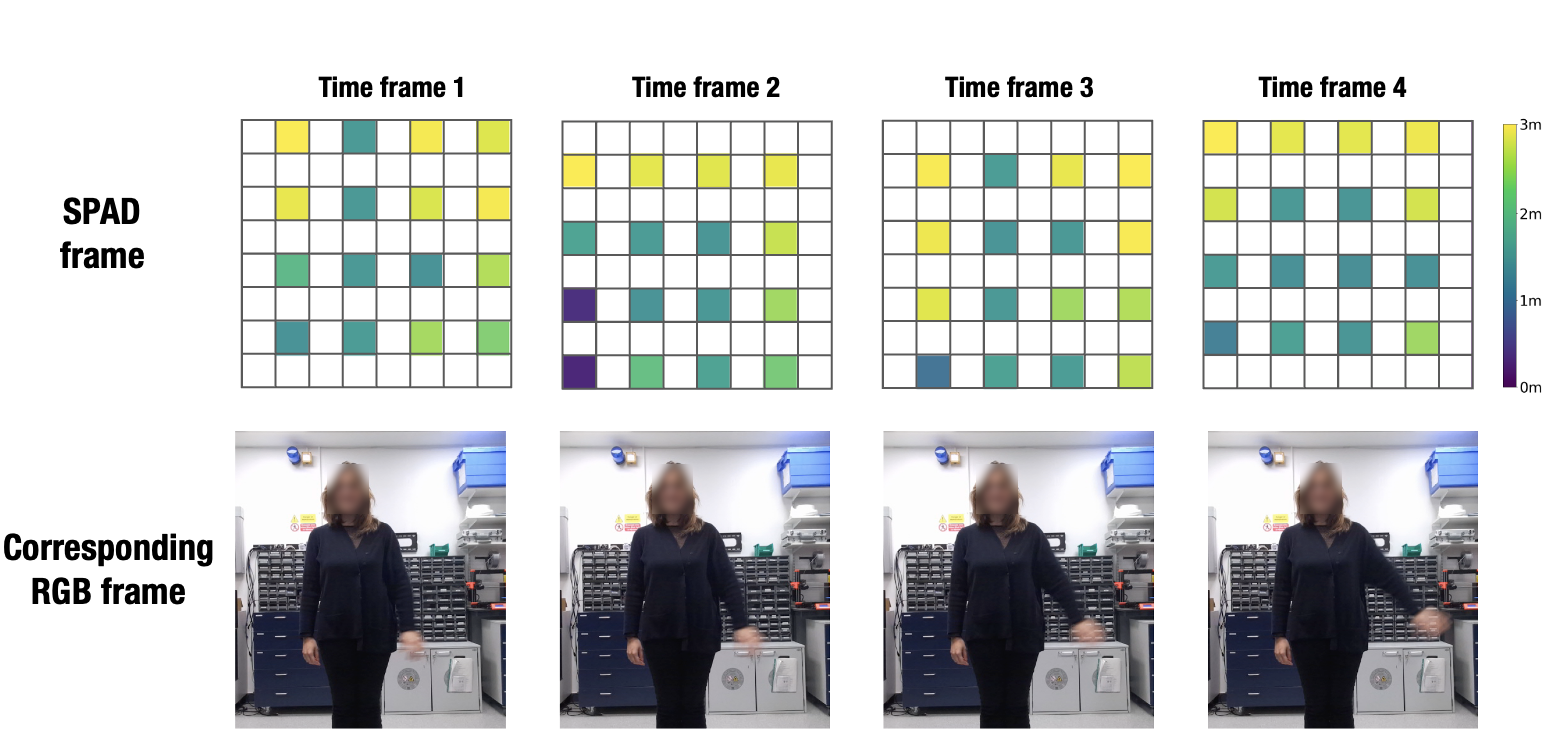}
\captionsetup{width=0.95\textwidth}
\caption{\textbf{Acquisition process of the STMicroelectronics sensor.} 
The ST sensor in 8x8 mode captures four consecutive snapshots of scattered pixels, combined into an 8x8 photon count histogram.
Motion in the scene can cause variations as pixels are captured at different times.  The figure illustrates the depth measurements (first row), with each pixel positioned according to the activated SPADs, alongside the corresponding synchronized RGB images obtained by a Kinect camera (second row). For simplicity, the depth is calculated based on the position of the maximum photon return from the histogram.
}
\label{fig:ST_acq}
\end{figure}

\begin{figure}[h!]
\centering
\begin{subfigure}{0.31\textwidth}
    \includegraphics[width=\textwidth]{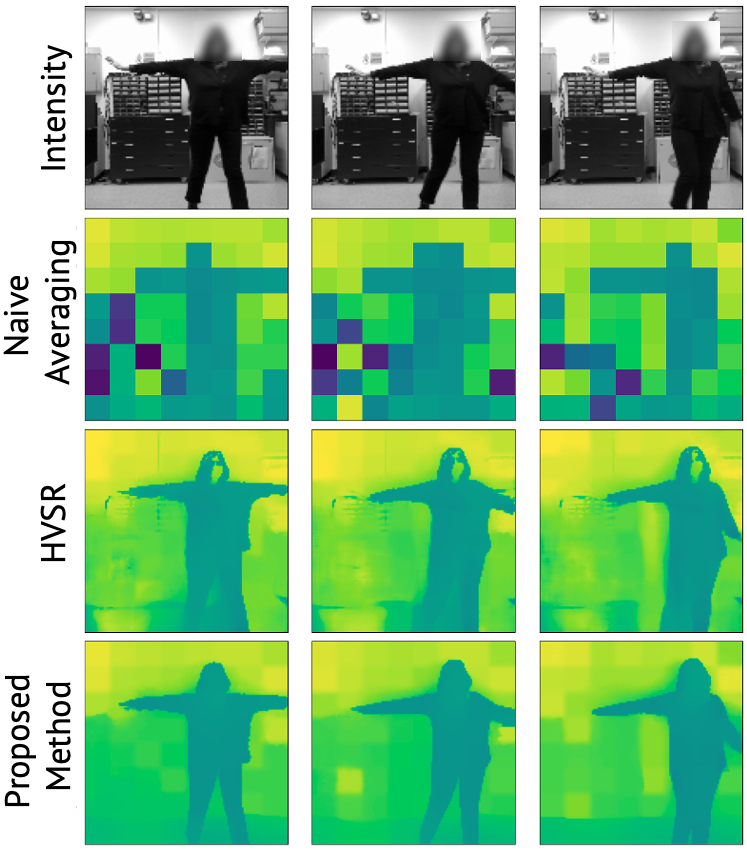}
\end{subfigure}
\hfill
\begin{subfigure}{0.282\textwidth}
    \includegraphics[width=\textwidth]{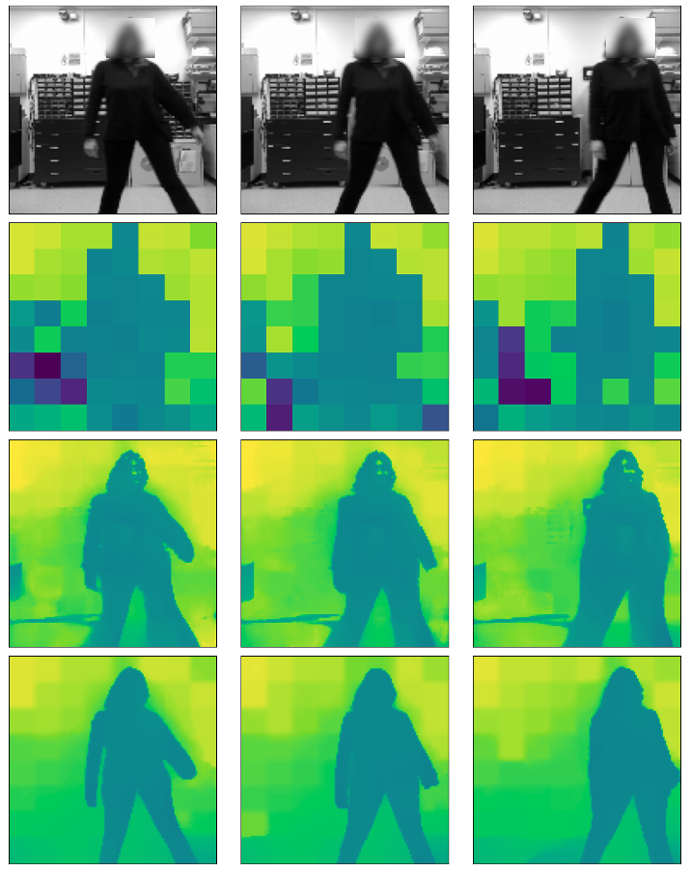}
\end{subfigure}
\hfill
\begin{subfigure}{0.282\textwidth}
    \includegraphics[width=\textwidth]{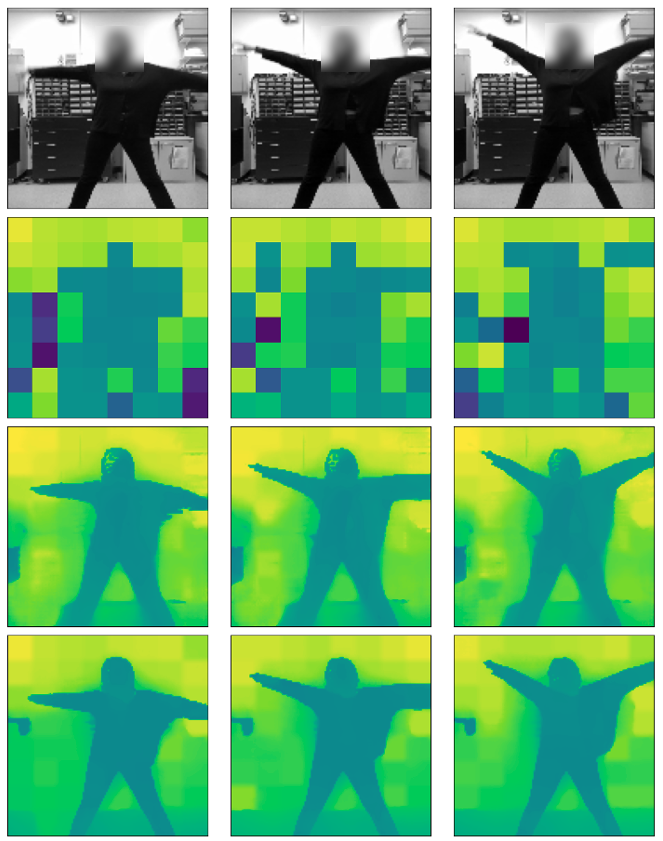}
\end{subfigure}
\begin{subfigure}{0.041\textwidth}
    \includegraphics[width=\textwidth]{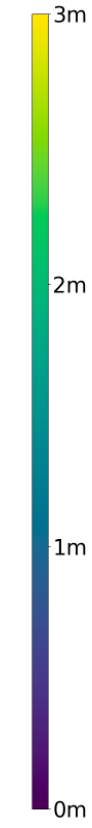}
\end{subfigure}
\captionsetup{width=0.95\textwidth}
\caption{\textbf{Scene images reconstructed from real measurements using a SPAD sensor from STMicroelectronics.} Each column represents a different time frame. The first row shows the intensity images captured by a Kinect camera. The second row displays the results obtained using the naive averaging approach. The third row presents the reconstruction results from the HVSR method \cite{sun2023consistent}. Finally, the fourth row demonstrates the results achieved with the proposed approach. A video representation of this scenario is provided in \blue{Visualisation 3}. }
\label{fig:ST_res}
\end{figure}

\vspace{-0.3cm}
\subsubsection{Reconstruction of people in motion using a cost-effective 8x8 pixel SPAD sensor in an indoor environment.}
The VL53L8 SPAD sensor from STMicroelectronics was used to demonstrate the super-resolution performance of our method on a real dataset. The system is paired with a Kinect camera to capture the high-resolution intensity images of the scene. The details of the SPAD sensor and of the Kinect camera are in Table \ref{table_systems}. The SPAD sensor records photon counts at a spatial resolution of $8\times8$ pixels. However, it does not capture all 64 pixels of the $8\times8$ grid in one snapshot. Instead,   it captures four sequential snapshots, each capturing 16 pixels scattered across a 4x4 uniform grid, see Figure \ref{fig:ST_acq}. As a result, the histograms $h_j$ at each time step are incomplete, with data recorded for only a portion of the pixels. To address this, we fill in the missing pixel values with zeros. Our reconstruction method is then applied to these zero-padded $8\times8$ histograms, along with the corresponding high-resolution RGB images, converted into intensity images $\bsR^t$, captured by the Kinect camera. \red{For this experiment, we set $M=4$ since four sequential snapshots are needed to obtain data for all $8 \times 8$ pixels. Accordingly, we adjust the Kinect's frame rate to be four times slower than the SPAD sensor’s frame rate, ensuring that a complete $8 \times 8$ SPAD frame is acquired between two consecutive intensity frames. }

Figure \ref{fig:ST_res} presents the reconstructions of the data using Naive Averaging, HVSR, and the proposed method for a scene of a moving person with multiple objects in the background. Our method yields the best results, producing more accurate shapes and effectively addressing the artifacts (e.g., facial features) generated by HVSR.

The method processes six frames simultaneously over five iterations, with a processing time of 23 seconds per iteration. Each iteration consists of two steps: the motion estimation and alignment process takes 15 seconds, and the super-resolution network HVSR \cite{sun2023consistent}, which runs on an NVIDIA RTX 6000 GPU, takes 8 seconds.

\vspace{-0.3cm}
\begin{figure}[h!]
\centering
\begin{subfigure}{0.45\textwidth}
    \includegraphics[width=\textwidth]{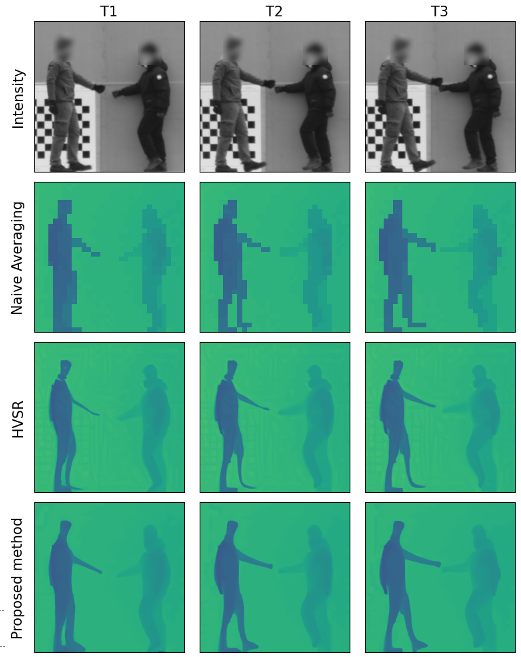}
    \subcaption{Two people in the scene}
\end{subfigure}
\hfill
\begin{subfigure}{0.45\textwidth}
    \includegraphics[width=\textwidth]{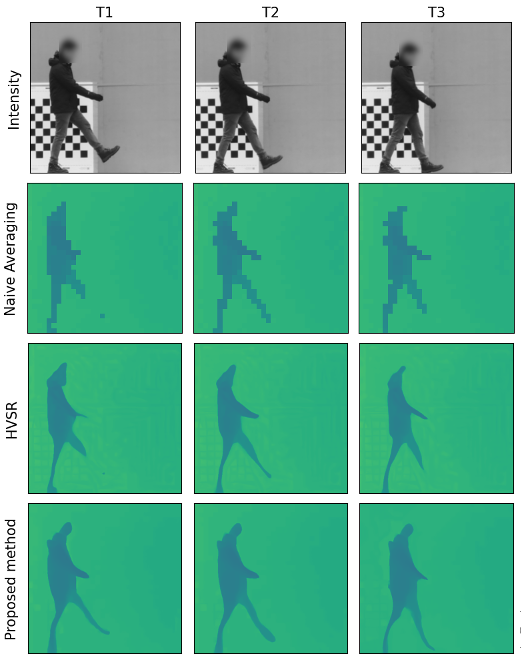}
    \subcaption{One person in the scene}
\end{subfigure}
\hfill
\begin{subfigure}{0.08\textwidth}
    \includegraphics[width=\textwidth]{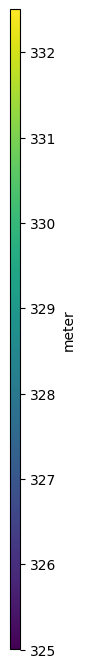} 
\end{subfigure}
\captionsetup{width=0.9\textwidth}
\caption{\textbf{Outdoor scene images reconstructed from measurements obtained with a SWIR SPAD LiDAR camera at 325m.} The columns represent different time frames. The scene of the left and right figures show two and one moving persons, respectively. Row 1 shows the intensity images; Row 2 shows the result of the naive averaging approach; Row 3 the results of HVSR \cite{sun2023consistent}; Row 4 displays results obtained using the proposed method. Video representations of these scenarios are provided in \blue{Visualisations 4 and 5}. }
\label{fig:300m}
\end{figure}

\begin{figure}[h!]
\centering
\begin{subfigure}{0.9\textwidth}
\includegraphics[width=\textwidth]{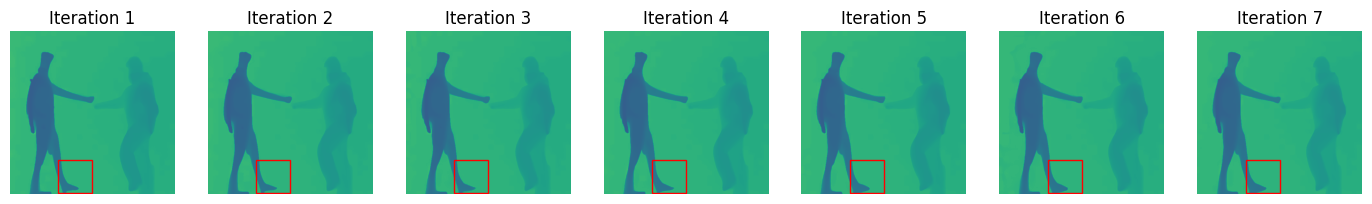}
\caption{Full field of view}
\end{subfigure}
\hfill
\begin{subfigure}{0.08\textwidth}
    \raisebox{2mm}{\includegraphics[width=\textwidth]{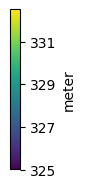}}
\end{subfigure}

\hfill
\begin{subfigure}{0.9\textwidth}
\includegraphics[width=\textwidth]{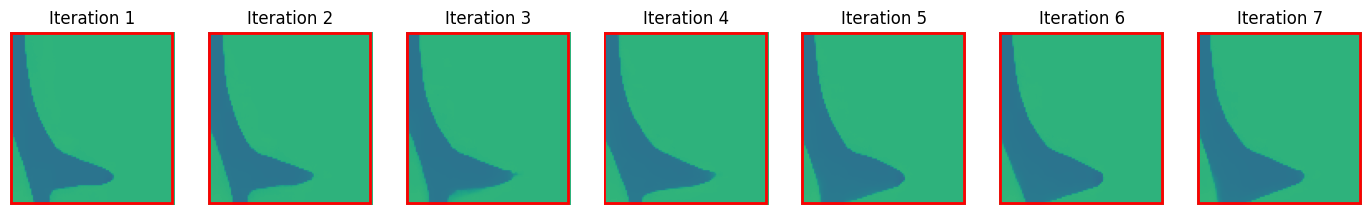}
\caption{Magnified views of the leg/foot of the person on lefthand side of the full field of view images above.}
\end{subfigure}
\hfill
\begin{subfigure}{0.08\textwidth}
    
    \raisebox{2mm}{\includegraphics[width=\textwidth]{Figures/Figure_10b.png} }
\end{subfigure}
\captionsetup{width=0.9\textwidth}
\caption{\textbf{Construction of scene image across multiple iterations of real outdoor measurements  acquired using a SWIR SPAD LiDAR camera located at a standoff distance of 325 meters.} The columns represent different iterations. First row show the depth maps, and second row the zoomed-in maps. }
\label{fig:PnP}
\end{figure}

\vspace{-0.3cm}
\subsubsection{Reconstruction of people in motion at a 325m range in an outdoor environment."} \vspace{-0.05cm}
\red{We use a short wave infrared (SWIR) InGaAs/InP SPAD camera operating at 1550-nm wavelength (Princeton Lightwave Kestrel) to perform outdoor experiments at long range  \cite{tachella2019real}. Using SWIR wavelengths in LiDAR  offers several advantages over shorter wavelengths: reduced solar background, better transmission through certain obscurants, and increased laser eye safety thresholds allowing safe use of higher laser power levels. These benefits can increase the maximum LiDAR range and/or improve depth quality with shorter acquisition times. This SPAD camera provides a stream of time-stamped binary frames at a rate of $150,400$ Hz. The system is coupled with a Canon camera that operates at a lower frame rate of $25$ Hz. 
To reduce computational load, we aggregate the SPAD camera's binary frames into histograms by performing naive averaging on groups of $1000$ frames before applying our algorithm. 
The RGB frames captured by the Canon camera are converted to intensity images and upsampled to $512 \times 512$, which provides a resolution 16 times higher than the binary frames, as required by HVSR (see section \ref{section_HVSR}). }
\red{In this experiment, the number of LiDAR frames acquired between two consecutive high-resolution intensity frames is $M=7$.
However, to simulate higher motion scenarios, we increase the parameter $M$ by skipping one or two intensity frames to include more LiDAR frames and to have larger motion changes in the scene. We choose $M=13$ in the case of the two-people dataset and  $M=37$ in the case of the one-people dataset. }

Figure \ref{fig:300m} shows reconstructions of two outdoor scenes with one and two people walking at a standoff distance of $325$ m from the sensor, using our proposed method, Naive Averaging, and HVSR \cite{sun2023consistent}.  We see that our method leads to more accurate images with complete limbs and sharper edges. Figure \ref{fig:PnP} illustrates the reconstruction process over multiple iterations using the proposed plug-and-play method. The figure demonstrates how the reconstruction improves progressively with each iteration. For example, as the iterations proceed, the details of the foot become increasingly clear, highlighting the effectiveness of the method in refining the reconstruction over time.
The method processes six frames simultaneously over six iterations, with a total processing time of 50 seconds per iteration. Each iteration consists of two steps: the super-resolution network HVSR \cite{sun2023consistent}, which takes 13 seconds, and motion estimation and averaging, which takes the remaining 37 seconds. 

\vspace{-0.05cm}
\section{Conclusion} \vspace{-0.05cm}
This paper has presented a novel plug-and-play method for 3D video super-resolution of single-photon data. The method effectively addresses motion blur and improves spatial resolution using a combination of optical flow and guided video super-resolution. The algorithm has been shown to work well across various noise levels, photon counts, pixel resolutions, sensor frame-rates and motion speeds. It was validated both on simulated and real-world data, demonstrating robustness in indoor and outdoor scenarios. We considered three practical cases using three SPAD sensors: (i) imaging very fast-moving objects with a hybrid sensor that alternates between intensity and histogram measurements at over 1000 fps, (ii) imaging with a few pixels using a cost-effective, compact, and lightweight sensor that can be easily integrated into consumer electronics such as smartphones, and (iii) imaging moving objects at long range with a short-wave infrared LiDAR, which reduces solar background noise and maintain eye safety at higher power levels compared to visible sensors, making it suitable for autonomous and self-driving vehicle applications. \red{
Our approach assumes aligned fields of view between the LiDAR and intensity cameras, which we plan to extend in future work. Additionally, we will explore the generation of inter-frame images to improve the temporal resolution of reconstructed 3D dynamic scenes.}

\begin{backmatter}
\bmsection{Acknowledgment}
We thank I. Gyongy, H.Li, S. Scholes, and J. Garcia Armenta, for their help in  collecting the data. 
\end{backmatter}

\begin{backmatter}
\bmsection{Funding}
This work was supported by the UK Royal Academy of Engineering under the Research Fellowship Scheme (RF/201718/17128), Serapis project C84 ST2   and EPSRC Grants EP/T00097X/1, EP/S026428/1.
\end{backmatter}

\begin{backmatter}
\bmsection{Code availability}
The proposed algorithm will be made available upon publication. 
\end{backmatter}

\begin{backmatter}
\bmsection{Disclosures}
The authors declare no conflicts of interest.
\end{backmatter}


\vspace{-0.15cm}

\bibliography{biblio}
\end{document}